\documentclass[twocolumn]{aastex63}
\usepackage{amsmath}
\usepackage{multirow}

\usepackage{ulem}

\newcommand{\NeII}{[Ne~{\sc ii}]}
\newcommand{\FeII}{[Fe~{\sc ii}]}
\newcommand{\NI}{[\ion{N}{1}]}
\newcommand{\OI}{[O~{\sc i}]}
\newcommand{\SII}{[\ion{S}{2}]}

\received{2022 July 23}
\revised{2024 February 6}
\accepted{2024 February 9}

\shorttitle{[NI]10400/10410 \AA{} Lines as Possible
Disk Wind Tracers} \shortauthors{Katoh et al. 2023}

\begin{document}

\title{\NI{} 10400/10410 \AA{} Lines as Possible Disk Wind Tracers in
  a Young Intermediate-Mass Star}

\correspondingauthor{Chikako Yasui}
\email{ck.yasui@gmail.com}

\author{Haruki Katoh}
\affiliation{Department of Physics, Graduate School of Science, Kyoto Sangyo University, Motoyama, Kamigamo, Kita-Ku, Kyoto, 603-8555, Japan}
\affiliation{Laboratory of Infrared High-resolution spectroscopy (LIH), Koyama Astronomical Observatory, Kyoto Sangyo University, Motoyama, Kamigamo, Kita-ku, Kyoto 603-8555, Japan}

\author{Chikako Yasui}
\affiliation{National Astronomical Observatory of Japan, 2-21-1 Osawa, Mitaka, Tokyo 181-8588, Japan}

\author{Yuji Ikeda}
\affiliation{Photocoding, 460-102 Iwakura-Nakamachi, Sakyo-ku, Kyoto 606-0025, Japan}
\affiliation{Laboratory of Infrared High-resolution spectroscopy (LIH), Koyama Astronomical Observatory, Kyoto Sangyo University, Motoyama, Kamigamo, Kita-ku, Kyoto 603-8555, Japan}

\author{Naoto Kobayashi}
\affiliation{Kiso Observatory, Institute of Astronomy, School of Science, The University of Tokyo, 10762-30 Mitake, Kiso-machi, Kiso-gun, Nagano 397-0101, Japan}
\affiliation{Institute of Astronomy, The University of Tokyo, 2-21-1 Osawa, Mitaka, Tokyo 181-0015, Japan}
\affiliation{Laboratory of Infrared High-resolution spectroscopy (LIH), Koyama Astronomical Observatory, Kyoto Sangyo University, Motoyama, Kamigamo, Kita-ku, Kyoto 603-8555, Japan}

\author{Noriyuki Matsunaga}
\affiliation{Department of Astronomy, Graduate School of Science,
University of Tokyo, Bunkyo-ku, Tokyo 113-0033, Japan}
\affiliation{Laboratory of Infrared High-resolution spectroscopy (LIH), Koyama Astronomical Observatory, Kyoto Sangyo University, Motoyama, Kamigamo, Kita-ku, Kyoto 603-8555, Japan}

\author{Sohei Kondo}
\affiliation{Kiso Observatory, Institute of Astronomy, School of Science, The University of Tokyo, 10762-30 Mitake, Kiso-machi, Kiso-gun, Nagano 397-0101, Japan}
\affiliation{Laboratory of Infrared High-resolution spectroscopy (LIH), Koyama Astronomical Observatory, Kyoto Sangyo University, Motoyama, Kamigamo, Kita-ku, Kyoto 603-8555, Japan}

\author{Hiroaki Sameshima}
\affiliation{Institute of Astronomy, The University of Tokyo, 2-21-1 Osawa, Mitaka, Tokyo 181-0015, Japan}
\affiliation{Laboratory of Infrared High-resolution spectroscopy (LIH), Koyama Astronomical Observatory, Kyoto Sangyo University, Motoyama, Kamigamo, Kita-ku, Kyoto 603-8555, Japan}

\author{Satoshi Hamano}
\affiliation{National Astronomical Observatory of Japan, 2-21-1 Osawa, Mitaka, Tokyo 181-8588, Japan}
\affiliation{Laboratory of Infrared High-resolution spectroscopy (LIH), Koyama Astronomical Observatory, Kyoto Sangyo University, Motoyama, Kamigamo, Kita-ku, Kyoto 603-8555, Japan}

\author{Misaki Mizumoto}
\affiliation{Science Education Research Unit, University of Teacher Education Fukuoka, 1-1 Akama-bunkyo-machi, Munakata, Fukuoka 811-4192, Japan}

\author{Hideyo Kawakita}
\affiliation{Laboratory of Infrared High-resolution spectroscopy (LIH), Koyama Astronomical Observatory, Kyoto Sangyo University, Motoyama, Kamigamo, Kita-ku, Kyoto 603-8555, Japan}
\affiliation{Department of Physics, Graduate School of Science, Kyoto Sangyo University, Motoyama, Kamigamo, Kita-Ku, Kyoto, 603-8555, Japan}

\author{Kei Fukue}
\affiliation{Laboratory of Infrared High-resolution spectroscopy (LIH), Koyama Astronomical Observatory, Kyoto Sangyo University, Motoyama, Kamigamo, Kita-ku, Kyoto 603-8555, Japan}
\affiliation{Education Center for Medicine and Nursing, Shiga University of Medical Science, Seta Tsukinowa-cho, Otsu, Shiga, 520-2192, Japan}

\author{Shogo Otsubo}
\affiliation{Laboratory of Infrared High-resolution spectroscopy (LIH), Koyama Astronomical Observatory, Kyoto Sangyo University, Motoyama, Kamigamo, Kita-ku, Kyoto 603-8555, Japan}

\author{Keiichi Takenaka}
\affiliation{Laboratory of Infrared High-resolution spectroscopy (LIH), Koyama Astronomical Observatory, Kyoto Sangyo University, Motoyama, Kamigamo, Kita-ku, Kyoto 603-8555, Japan}
\affiliation{Department of Physics, Graduate School of Science, Kyoto Sangyo University, Motoyama, Kamigamo, Kita-Ku, Kyoto, 603-8555, Japan}




\begin{abstract}

In this study, we performed high-resolution near-infrared (NIR)
spectroscopy ($R=28$,000; $\lambda = 0.90$--1.35 $\mu$m) with a high
signal-to-noise ratio on HD 200775, a very young ($\sim$0.1 Myr old)
and massive intermediate-mass star (a binary star with a mass of about
10 $M_\odot$ each) with a protoplanetary disk.
The obtained spectra show eight forbidden lines of three elements: two
of \SII{} (10289 and 10323 \AA), two of \NI{} (10400 and 10410 \AA), and
four of \FeII{} (12570, 12946, 12981, and 13209 \AA).
This is the first time that the \NI{} lines are detected in a young
stellar object with a doublet deblended.
Gaussian fitting of the spectra indicates that all line profiles have
low-velocity components and exhibit blueshifted features, suggesting
that all lines originate from the disk winds (magnetohydrodynamic disk
wind and/or photoevaporative wind).
Based on the fit, the \NI{} and \FeII{} lines are categorized into
narrow components, while the \SII{} lines are at the boundary between
broad and narrow components.
These forbidden lines are suggested to be very promising
disk wind tracers among the existing ones because
they are in the NIR-wavelength range, which can be observed from early
stages with high sensitivities.
Among these lines, \NI{} lines would be a rather powerful probe for
deriving the basic physical parameters of disk wind gases.
However, the study of these lines herein is limited to one object; thus,
further studies are needed to examine their properties.

\end{abstract}



\section{Introduction} \label{sec:Intro}

Protoplanetary disks are disks of gas and dust surrounding a young star 
that are formed when the star is born in a molecular cloud and then
gradually disappear as the central star evolves into a main-sequence
star.
Because planets are formed in the process of disk evolution/dispersal,
understanding the mechanism of the process is essential to elucidate the
planet formation process.
Disk dispersal can be divided into two main processes: mass accretion
\citep{{Shakura1973},{Lynden-Bell1974}}, in which disk gas accretes to
the central star, and disk dissipation, in which disk gas is ejected
into interstellar space.
Mass accretion activities have been confirmed in many objects, mainly by
the observation of hydrogen emission lines.
Comparisons of the fluxes and line profiles with theoretical models
have revealed the basic properties of mass accretion, such as the mass
accretion rate and the geometric structure of the accretion, and
enabled their quantitative derivation \citep{Hartmann2016}.

Disk winds, such as magnetohydrodynamic (MHD) disk winds (e.g.,
\citealt{Suzuki2009}) and photoevaporative disk winds (e.g.,
\citealt{Hollenbach+1994}), are considered to be the dominant
mechanisms of disk dissipation.
Recently, an increasing number of atomic and molecular transitions
with line profiles indicative of disk winds have been observed (e.g.,
\citealt{Pascucci2023}), including optical forbidden lines (e.g.,
\OI{} 5577/6300 \AA{} and \SII{} 4068/4076 \AA{};
\citealt{Hartigan1995}), near-infrared (NIR) CO and H$_2$ transitions
\citep{Bary2003}, infrared forbidden lines (e.g, \NeII{} 12.8 $\mu$m;
\citealt{Pascucci2009}, \citealt{Pascucci2020}), and CO rotational
lines at millimeter wavelengths \citep{Trapman2022}.
Among them, forbidden lines with blueshifted emission features will be
the focus of this paper.

The line profiles of forbidden lines often show multiple components,
including high-velocity components (HVCs; $|V_c| \approx 50$--300 km
s$^{-1}$), originating from jets,
and low-velocity components (LVCs) with a central velocity $|V_c| \le
30$ km s$^{-1}$, which originate from disk winds (e.g.,
\citealt{Hartigan1995}).
Recent high-dispersion spectroscopic observations have revealed that
LVCs often have two components: a narrow component (NC) with a
velocity width in FWHM $\Delta V_{\rm FWHM} \le 40$ km s$^{-1}$ and a
broad component (BC) with $\Delta V_{\rm FWHM} > 40$ km s$^{-1}$, 
although they cannot be clearly distinguished because they partially
overlap \citep{Simon+2016, Banzatti+2019}.
The complex profiles of LVCs with both BCs and NCs
are not fully explained by MHD or photoevaporative disk winds, 
and their origin remains under debate.
The BC in \OI{} lines is inferred to be launched from radii of
$\lesssim$0.5 au, ruling out a photoevaporative wind and instead
pointing to an origin in MHD disk winds \citep{Simon+2016}. Meanwhile,
the NC is launched at larger disk radii of $\gtrsim$1 au (as inferred
from the smaller line widths), and its origin remains contentious to
date.
For example, after correlating the BCs and NCs,
  \citet{Banzatti+2019} concludes that the luminosities decrease as
the inner disk is dissipated, suggesting that both components
originate in the MHD disk wind.
A spectroastrometric study \citep{Whelan2021} also suggests that
the NC traces MHD disk wind.

Models of the disk winds, including detailed heating and cooling
mechanisms, have been published and compared with the observed
profiles of the forbidden lines from the observations, while for the
molecular tracers there is currently a lack of a suitable hydrodynamic
wind model coupled with a chemical model and a dust evolution model,
and the interpretation of the observations has not progressed
\citep{{Pascucci2023}, {Ercolano&Pascucci2017}}.
\NeII{} 12.8 $\mu$m model line profiles were produced based on
photoevaporative winds (e.g, \citealt{Alexander2008},
\citealt{Ercolano&Owen2010}) and appear to reproduce well the
profiles obtained from observations \citep{Pascucci2020}.
Model line profiles of optical forbidden lines have also been
calculated, for example, in \citet{Weber2020}, suggesting that the
combination of MHD and optical evaporation winds can explain the BC
and NC composite profiles.

In this paper, we explored disk wind tracers in the NIR-wavelength
range because they are more sensitive than observations in the
optical- and mid-IR- (MIR-) wavelength ranges, where the disk wind
tracers identified so far are located.
In general, young objects that are still embedded in molecular clouds
are difficult to observe owing to large extinctions in the optical
bands and thermal radiations from the telluric atmosphere in the MIR
range.
Herein, we focus on intermediate-mass stars ($\sim$2--10 $M_\odot$),
which are more massive than low-mass stars that have been the subject
of previous observations.
We investigated novel disk wind tracers in the NIR-wavelength region,
where emission lines from young stars are easily detected with a high
signal-to-noise ratio (S/N).
For example, several theoretical models have indicated that the higher
photoevaporation efficiency of intermediate-mass stars
\citep[e.g.,][]{Gorti+2009,Komaki2021,Kunitomo+2021} can explain why
the disk lifetime of these stars is shorter than that of low-mass
stars \citep{Yasui+2014,Ribas+2014}, and direct observation of
intermediate-mass stars would be an important next step.
As a first target, we selected HD 200775, which is a young stellar
object (YSO) with a massive intermediate stellar mass.
The disk of HD 200775 is directly imaged in MIR bands
\citep{Okamoto+2009}.

The rest of this paper is organized as follows.
Section~2 summarizes the properties of the target, observation, and data
reduction.
Section~3 presents the obtained NIR spectra, identifies the
forbidden lines, and discusses their properties.
In Section~4, we categorize the line profiles of the forbidden lines,
discuss their origins and the possibility of disk wind in
protoplanetary disks around intermediate-mass stars, and use the NIR
forbidden lines detected in this study as new disk wind tracers.
Finally, the paper concludes in Section~5.


\section{Target, Observation, and Data Reduction} \label{sec:Obs}

\subsection{HD 200775} \label{subsec:HD200775}

We selected HD 200775 as a target, which has the earliest spectral type
among YSOs with confirmed protoplanetary disks from the Catalog of
Circumstellar
Disks\footnote{\url{https://www.circumstellardisks.org/index.php}}.
Optical spectroscopy has shown that the spectral type is B3
\citep[e.g.,][]{Hernandez2004}.
From Gaia Data Release 3 \citep{GaiaDR3}, its astrometric distance is
estimated as $348^{+5}_{-4}$ pc.
HD 200775 is a binary system based on the time variability of the H$\alpha$
emission line \citep{Miroshnichenko+1998}. 
The primary and secondary masses are $10.7\pm2.5$ $M_\odot$ and $9.3 \pm
2.1$ $M_\odot$, respectively.
The semimajor axis of the orbit is 5.4 au, assuming a distance of 350 pc
\citep{Monnier+2006,Alecian+2008}.
The object is estimated to be very young ($\sim$0.1 Myr)
\citep{Alecian+2008}, and the mass accretion rate ($\dot{M}_{\rm acc}$)
and mass-loss rate ($\dot{M}_{\rm loss}$) are very high
($\dot{M}_{\rm acc} = 10^{-4.2}$ $M_\odot$ yr$^{-1}$ and $\dot{M}_{\rm
loss} = 10^{-6.55}$ $M_\odot$ yr$^{-1}$)
\citep{Arun+2019,Damiani+1994}.

Considering the possibility of disk wind in HD 200775,
\citet{Okamoto+2009} reported that the free-free emission detected by
\citet{Fuente+2001} may be generated from the photoevaporation region
in this star (although \citet{Fuente+2001} considered the emission as
a stellar wind origin) because $R_g$ calculated based on the stellar
mass (73--146 au) agrees well with the hole size of the disk obtained
from their direct imaging in MIR bands.
From the imaging, the disk inclination angle is estimated as $i =
54\fdg1 \pm 1\fdg2$.

\subsection{Observation} \label{subsec:Obs}

We obtained the spectra of HD 200775 using the NIR spectrograph,
WINERED \citep{Ikeda2022}, which is attached to the Araki 1.3 m
telescope at Koyama Astronomical Observatory, Kyoto Sangyo University,
Japan \citep{Yoshikawa+2012}.
WINERED is a high-sensitivity and high-resolution spectrograph.
It provides two resolution modes with spectral resolving powers $R$ of
28,000 and 70,000, called the ``WIDE'' and ``HIRES'' modes,
respectively, and is customized for short NIR bands at 0.9--1.35 $\mu$m.
In the WIDE mode, which was used in this study, the entire wavelength
range of 0.9--1.35 $\mu$m is covered simultaneously in a single
exposure. WINERED has a 1.7 $\mu$m cutoff $2048 \times 2048$ HAWAII-2RG
infrared array with a pixel scale of $0\farcs75$ pixel$^{-1}$.
We used a slit of $1\farcs65$ wide and 45$\arcsec$ long.
Observations were performed on 2015 July 20. The sky condition was
photometric with a relative humidity of 60\%--75\% throughout the
night.
The seeing size was 3$\arcsec$--4$\arcsec$.
To achieve an accurate sky subtraction, all the observations were made
by nodding the target along the slits on two slit positions, ``A'' and
``B,'' with one exposure per position.
The spectra of ``B'' (``A'') are used for the sky spectra of ``A''
(``B'').
This method is called ABBA dithering.
We took four exposures (two A positions and two B positions), and the
exposure time per frame was 600 s, resulting in a total exposure time
of 2400 s for the target.
$\rho$ Peg with a spectral-type A1V was selected as the telluric
standard star.
It was considered with ABBA nodding within 1 hr of target data acquisition.
The exposure time for a single frame was set to 120 s, and four sets of
data were obtained, resulting in a total exposure time of 480 s.
Details of the observations are listed in Table~\ref{tab:ObsSummary}.

\subsection{Data Reduction} \label{subsec:reduction}

All data were reduced using a WINERED data-reduction pipeline
\citep{Hamano2024}.
This pipeline includes the standard procedures of echelle
spectroscopy: bad pixel masking, sky subtraction, flat-fielding,
scattered light subtraction, transformation of two-dimensional (2D)
echelle image into a rectangular image with orthogonal space and
wavelength axes, spectrum extraction, and continuum normalization.
To extract the one-dimensional spectrum, the pixels corresponding to
twice the FWHM (7\arcsec) were integrated in the spatial direction.
The spectrum of HD 200775 was divided by the spectrum of the standard
star ($\rho$ Peg) in the procedure by \citet{Sameshima+2018}, which
eliminates the photospheric features to correct the telluric
absorption lines.
For each echelle order, the spectra were normalized to 1.0 using
the IRAF\footnote{IRAF is distributed by the National
Optical Astronomy Observatories, which are operated by the Association
of Universities for Research in Astronomy, Inc., under cooperative
agreement with the National Science Foundation.}/PyRAF continuum task.
Finally, the spectra were corrected for heliocentric radial velocities
(RVs) with 8.65 km s$^{-1}$, which varied within 0.01 km s$^{-1}$
throughout the observations, using the IRAF/PyRAF rvcorrect task.


\section{Results} \label{sec:Result}
\subsection{Detection of the Forbidden Lines} \label{subsec:fline}

From the HD 200775 spectra obtained in Section~\ref{subsec:reduction},
we focus on forbidden lines that can be disk wind tracers.
Forbidden lines originate only from regions with electron densities
lower than the critical electron density defined for each transition
and have little or no self-absorption due to their low transition
probabilities \citep{Osterbrock2006}.
Therefore, forbidden lines can trace low-electron-density regions
($n_{\rm e} \leq 10^{8}$ cm$^{-3}$), where disk wind occurs.
In addition, they can be very useful for accurately obtaining the
velocity fields from their profiles.

In the obtained spectra, eight forbidden lines of three elements, \SII,
\NI, and \FeII, were detected at $S/N \geq 10$:
two in \SII{} (10289 and 10323 \AA), two in \NI{} (10400 and 10410 \AA),
and four in \FeII{} (12570, 12946, 12981, and 13209 \AA).
All forbidden lines are multiplets.
When measuring the S/N of the emission lines, the noise levels were
estimated from a region 3 \AA{} away from both sides of the emission
lines for 2 \AA, a total of 4 \AA{} from the long and short wavelength
sides of the spectra.
In the case where other lines were present in the 2 \AA{} regions, the
other continuum regions were used, where noise levels can be estimated
in the same order as the emission lines.
The S/Ns were calculated from the integrated intensity in the FWHM of
the emission lines ($S$) and the measured noise level ($\sigma_N$) using
the formula ${\rm SNR} = S / \sqrt{\sigma_{N}^{2} \times N_{\rm FWHM}}$,
where $N_{\rm FWHM}$ is the number of pixels included within the FWHM.

\begin{figure*}[ht]
  \begin{center}
  \epsscale{1.15} \plotone{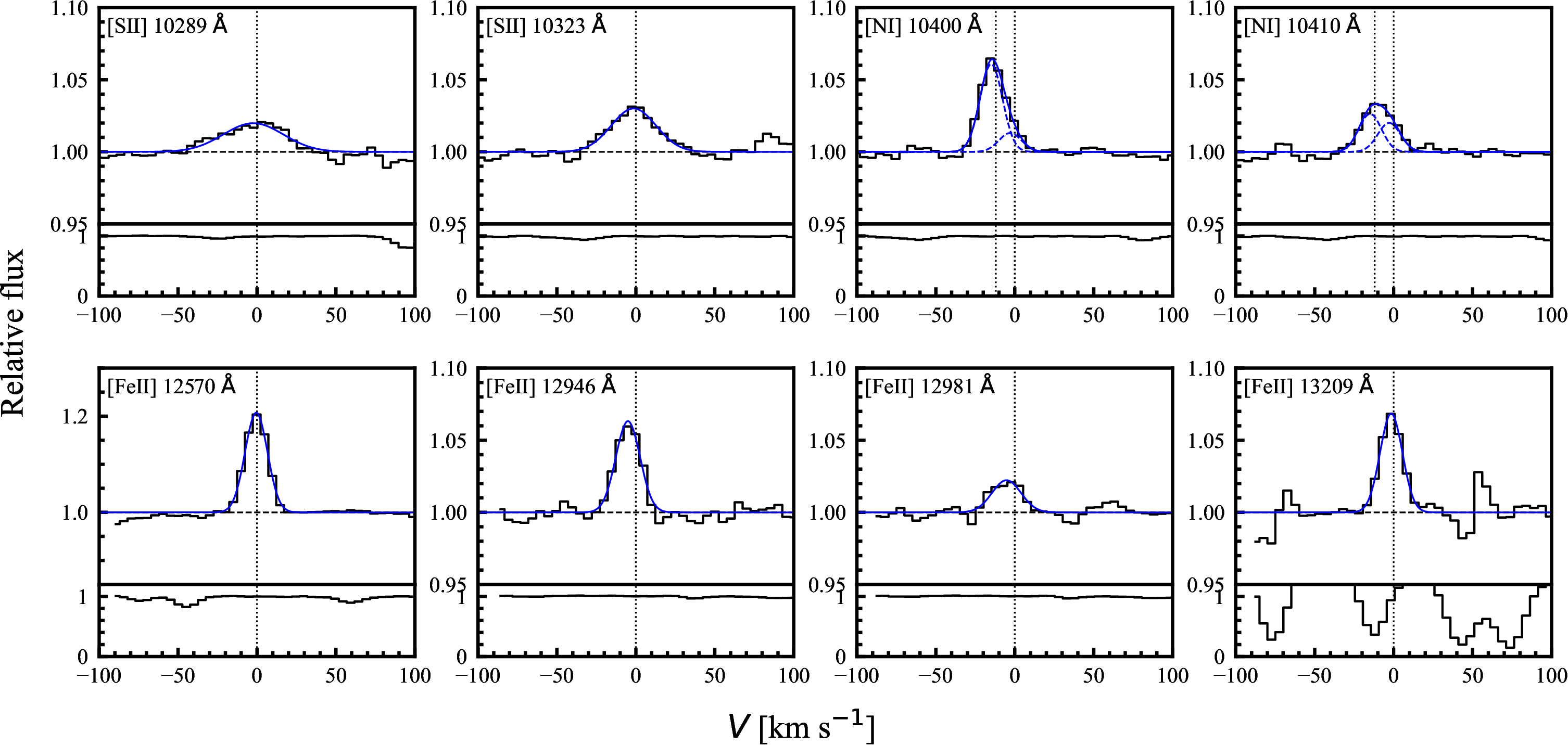}
 \caption{Line profiles of the forbidden lines detected in HD 200775
(upper panels).
The spectra obtained with WINERED are shown in black, and the fitting
results with the Gaussian profile are shown in blue.
For \NI{} lines, because each line is a doublet line, the obtained
spectrum is fitted with two components and shown by the blue dashed
line, and the composite of the two components is shown by the blue solid
line.
The velocities are relative to the stellar rest velocities. The
normalized spectra of the telluric standard star are shown in the lower
panels.}
\label{fig:spectrum}
\end{center}
\end{figure*}

The properties of the detected lines are summarized in
Table~\ref{tab:flinelist}.
The table lists the wavelength, Einstein $A$ coefficient, and transition
of each line obtained from NIST Atomic Spectra Database Lines
Data\footnote{\url{https://www.nist.gov/pml/atomic-spectra-database}}.
All detected lines are shown as black lines in Figure~\ref{fig:spectrum}
with respect to the stellar velocities.
Based on several derivations of RV (see \citealt{Bisyarina2015} for
references),
the measurements from stellar lines are expected to provide the most
reliable estimates of the RV correction. 
In addition, as HD 200775 is a binary system, the systemic velocities
derived from the data of two separated RVs over a long time period
(years) are considered to be the most reliable.
From this perspective, the derivations of \citet{Alecian+2008}
($-$7.9$\pm$0.9 km s$^{-1}$) and \citet{Bisyarina2015} ($-$9.7$\pm$2.6
km s$^{-1}$) are considered as the most reliable velocities, and their
weighted average, $-$8.4$\pm$1.0 km s$^{-1}$, was adopted in the
present analysis.

\subsection{Derivation of Velocity Fields for the Detected Forbidden Lines}
\label{subsec:flineproperty}

In this section, we determine the velocity fields for the forbidden
lines detected in the HD 200775 spectra, namely \SII, \NI, and \FeII.
The spectra are fitted to a Gaussian profile with three parameters,
the centroid velocity ($V_c$), the velocity width in FWHM ($\Delta
V_{\rm FWHM (obs)}$), and the peak intensity relative to the continuum
level ($I_{\rm peak}$).
For the \NI{} 10400/10410 \AA{} lines, because the emission lines are
intrinsically doublets and the profiles are slightly asymmetric, each
doublet was fitted to two components.
Because each emission line requires three parameters for fitting, a
naive fitting of the quadruple lines in \NI{} requires 12 parameters.
Assuming that collisional deexcitation is inefficient when forbidden
lines are emitted,
the intensity ratio of lines with a common upper level should be
determined only by the ratio of their Einstein $A$ coefficients.
Among the four \NI{} lines, the lines at 10401.004 and 10410.021
\AA\ have common upper levels of 2s$^2$\ 2p$^3$\ $^2$P$^{o}_{3/2}$,
while those at 10400.587 and 10410.439 \AA\ have common levels of
2s$^2$\ 2p$^3$\ $^2$P$^{o}_{1/2}$.
In addition, because the energy difference between the two upper
levels is very small, these four lines are assumed to have the same
regional origin, that is, the same $V_c$ and $\Delta V_{\rm FWHM
  (obs)}$.
Therefore, the original 12 fitting parameters can be reduced to four
(one $V_c$ value, one $\Delta V_{\rm FWHM (obs)}$ value, and two
$I_{\rm peak}$ values).

In Figure~\ref{fig:spectrum}, the blue and black lines are the
obtained fitting profiles and the HD 200775 spectra, respectively.
The intrinsic $\Delta V_{\rm FWHM}$ was calculated from the
instrumental profile correction on $\Delta V_{\rm FWHM (obs)}$ as
$\sqrt{\Delta V_{\rm FWHM (obs)}^2 - \Delta V_{\rm FWHM (inst)}^2}$,
where $\Delta V_{\rm FWHM (inst)}$ is the instrumental velocity width in
FWHM.
Because the spectral resolution of WINERED is almost constant ($R =
28$,000) over the entire wavelength range, the $\Delta V_{\rm FWHM
  (inst)}$ of all emission lines was assumed as 10.70 km s$^{-1}$.

Table~\ref{tab:flinelist} lists the three parameters, $V_c$, $\Delta
V_{\rm FWHM}$, and $I_{\rm peak}$, obtained from the fitting for each
emission line.
The uncertainties in $V_c$ were sourced from those in the systemic
velocity of HD 200775 ($\pm$1.0 km s$^{-1}$) and the observational
errors.
The $V_c$ values obtained from all detected emission lines were
$\gtrsim$$-$5 km s$^{-1}$, indicating slightly blueshifted features.
The \NI{} and \FeII{} lines show narrow line widths with $\Delta
V_{\rm FWHM}$ of $\sim$15--20 km s$^{-1}$, whereas the \SII{} lines
show slightly wider line widths with $\Delta V_{\rm FWHM}$ of
$\sim$30--40 km s$^{-1}$.


\section{Discussion} \label{sec:Discussion}

\subsection{Origin of the Forbidden Lines\label{subsec:origin}}

\begin{figure*}[ht]
  \begin{center}
  \epsscale{0.5} \plotone{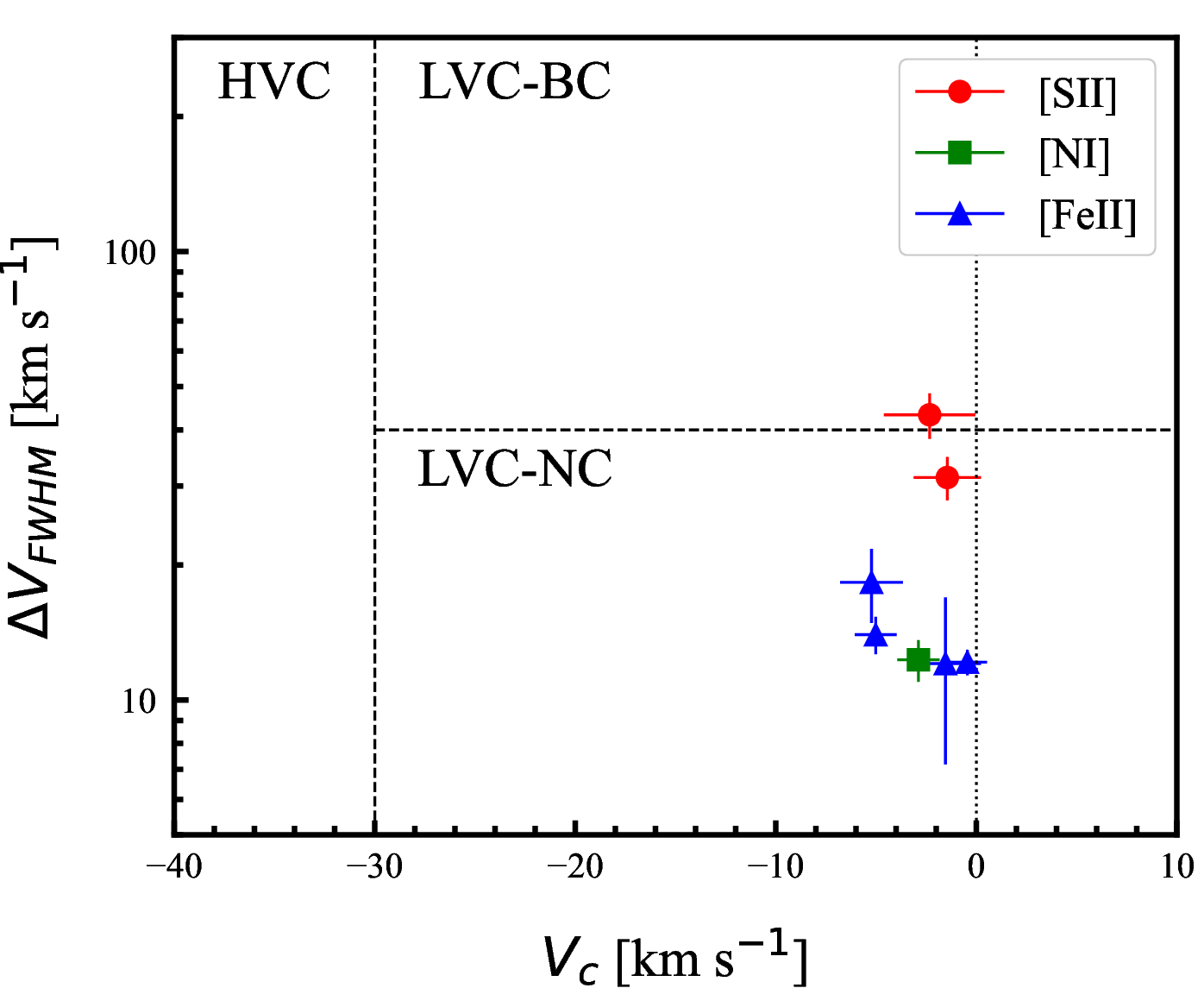}

\caption{Velocity properties ($V_c$ vs. $\Delta V_{\rm FWHM}$) of the
  detected forbidden lines in HD 200775.
The circles, squares, and triangles with error bars show \SII{}, \NI{},
and \FeII, respectively. 
The classifications divided by the dashed lines, HVC ($V_c \lesssim -30$
km s$^{-1}$),
LVC-BC ($V_c \gtrsim -30$ km s$^{-1}$ and $\Delta V_{\rm FWHM} \ge 40$
km s$^{-1}$), and LVC-NC ($V_c \gtrsim -30$ km s$^{-1}$ and $\Delta
V_{\rm FWHM} \le 40$ km s$^{-1}$), were adapted from
\citet{Simon+2016}.}  \label{fig:vplot}
\end{center}
\end{figure*}

Figure~\ref{fig:vplot} categorizes the velocity properties of the
forbidden lines into different velocity fields, HVC, LVC-BC, and
LVC-NC (\citealt{Simon+2016}; see Section~\ref{sec:Intro}). 
In this figure, the velocity properties of the detected forbidden
lines (\SII, \NI, and \FeII) are plotted as circles, squares, and
triangles, respectively.
The quadruple \NI{} lines are shown here as a single plot because
they should have the same $V_c$ and $V_{\rm FWHM (obs)}$ values in the
fitting (Section~\ref{subsec:flineproperty}).
The figure shows that $|V_c| \le 30$ km s$^{-1}$ for all emission
lines and that all lines were classified as LVC, suggesting that
they originated from disk wind.
The line profiles \NI{} and \FeII{} with narrow velocity widths
($\Delta V_{\rm FWHM} \simeq 10$--20 km s$^{-1}$) were categorized as
LVC-NC whereas those of \SII{} with slightly wider velocity widths
($\Delta V_{\rm FWHM} \sim 30$--40 km s$^{-1}$) were located at the
boundary between LVC-BC and LVC-NC.

Note that line centroids for some lines (two \SII{} lines and two
\FeII{} lines) are within 1$\sigma$ to the stellar velocity.
This may be because the disk inclination angle of HD 200775 is
relatively large ($i = 54\fdg1 \pm 1\fdg2$), which makes blueshifted
features of line profiles less noticeable \citep{Alexander2008}. 
Actually, disk wind tracers generally have small $V_c$ values and some
are estimated to be close to the stellar velocity or even redshifted
(e.g., \citealt{Banzatti+2019} for the \OI{} case).
Nevertheless, the results that all forbidden lines detected here are
estimated as blueshift suggest that they are all disk wind tracers.
In particular, the result that all four \FeII{} lines, which are close
to each other in energy level, show blueshifted features statistically
supports that they are indeed blueshifted.

In previous studies, the same lines were also detected in YSOs
(mainly in low-mass stars).
The \SII{}, \NI{}, and \FeII{} lines with HVCs have been detected in
ESO-H$\alpha$ 574 and Par-Lup 3-4 \citep{Bacciotti+2011,
  Giannini+2013, Whelan2014}.
These lines have also been detected in other T Tauri stars, and all
lines were classified in the literature as LVC ($-$40 $\lesssim v
\lesssim 40$ km s$^{-1}$), MVC (medium velocity component; $-$100
$\lesssim v \lesssim -40$ km s$^{-1}$), or HVC ($v \lesssim -100$ km
s$^{-1}$).
These lines with HVCs were also detected in a low-mass Herbig Halo
object, HH34 \citep{Nisini+2016}.
The previous studies discussed these lines as jet tracers.
The reason why these lines were detected as LVCs in HD 200775 but
observed as various components in the low-mass T Tauri stars is
unclear but may be explained by differences among the YSO masses. 
To resolve this problem, the \NI{} line profiles must be
comprehensively investigated through further WINERED observations or
analyses of archival data.
Moreover, to our knowledge, \NI{} lines with a deblended doublet have
not been detected in YSOs before.
Notably, the two \NI{} lines are further doublet (10400.587 and
10401.004 \AA{} for \NI{} 10400 \AA{}, and 10410.021 and 10410.439 \AA{}
for \NI{} 10410 \AA), making quadruple lines in total, but they are not
resolved due to their intrinsic widths (see the following subsection).

\subsection{Possible Origins Other than Stellar and Circumstellar Material Origin} \label{subsec:porigin}

The forbidden lines (\SII, \NI, and \FeII{}) detected in HD 200775
may not be intrinsic features of the object.
\citet{Lowe+1979}, \citet{Luhman+1998}, and \citet{Walmsley+2000}
detected \SII{} 10286/10320 \AA, \NI{} 10400/10410 \AA, and \FeII{}
12570, 12946, 12981, 13209 \AA{} forbidden lines in the bar structure in
the Orion Nebula and photodissociation regions (PDRs).
HD 200775 is located at the center of NGC 7023, a reflection nebula.
From imaging observations of ultraviolet and optical bands, which have
confirmed scattered light from the dust, the dust nebula extends to a
radius of approximately 5$\arcmin$
\citep[e.g.,][]{Witt&Cottrell1980,Witt+1992}.
Furthermore, the nearby region of HD 200775 ($\le$3$\arcmin$) is
surrounded by particularly dense nebular components.
Imaging of the H$_2$ (1.18 and 1.21 $\mu$m) and CO emission lines in
the NIR and radio wavelengths, respectively, has confirmed the
presence of a bar structure, which is a PDR, approximately 1$\arcmin$
away in the northwest and south directions from the central star
\citep[e.g.,][]{Sellgren+1992,Lemaire+1996,Gerin+1998,Witt+2006}.
Therefore, the forbidden lines detected here may not be intrinsic
features of HD 200775 but contaminations from the foreground nebulae
surrounding the objects.
It could also be the influence of OH emission lines from the terrestrial
atmospheric origin.
To evaluate these possibilities, we investigated the spectra of regions
somewhat away from the optical center of HD 200775 (hereafter,
``reference regions'').

\begin{figure*}[ht]
\begin{center}

\epsscale{0.75}\plotone{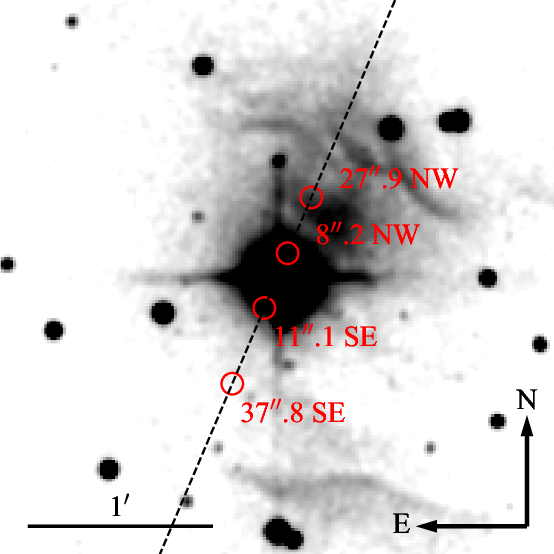}

\caption{Spatial map around HD 200775.  Red circles indicate the
  positions of the reference regions superposed on the Two Micron All
  Sky Survey $J$ band image.
The size of each circle is the seeing size.}
\label{fig:ref_region_map}
\end{center}
\end{figure*}

\begin{figure*}[ht]

  \begin{center}
\epsscale{0.75} \plotone{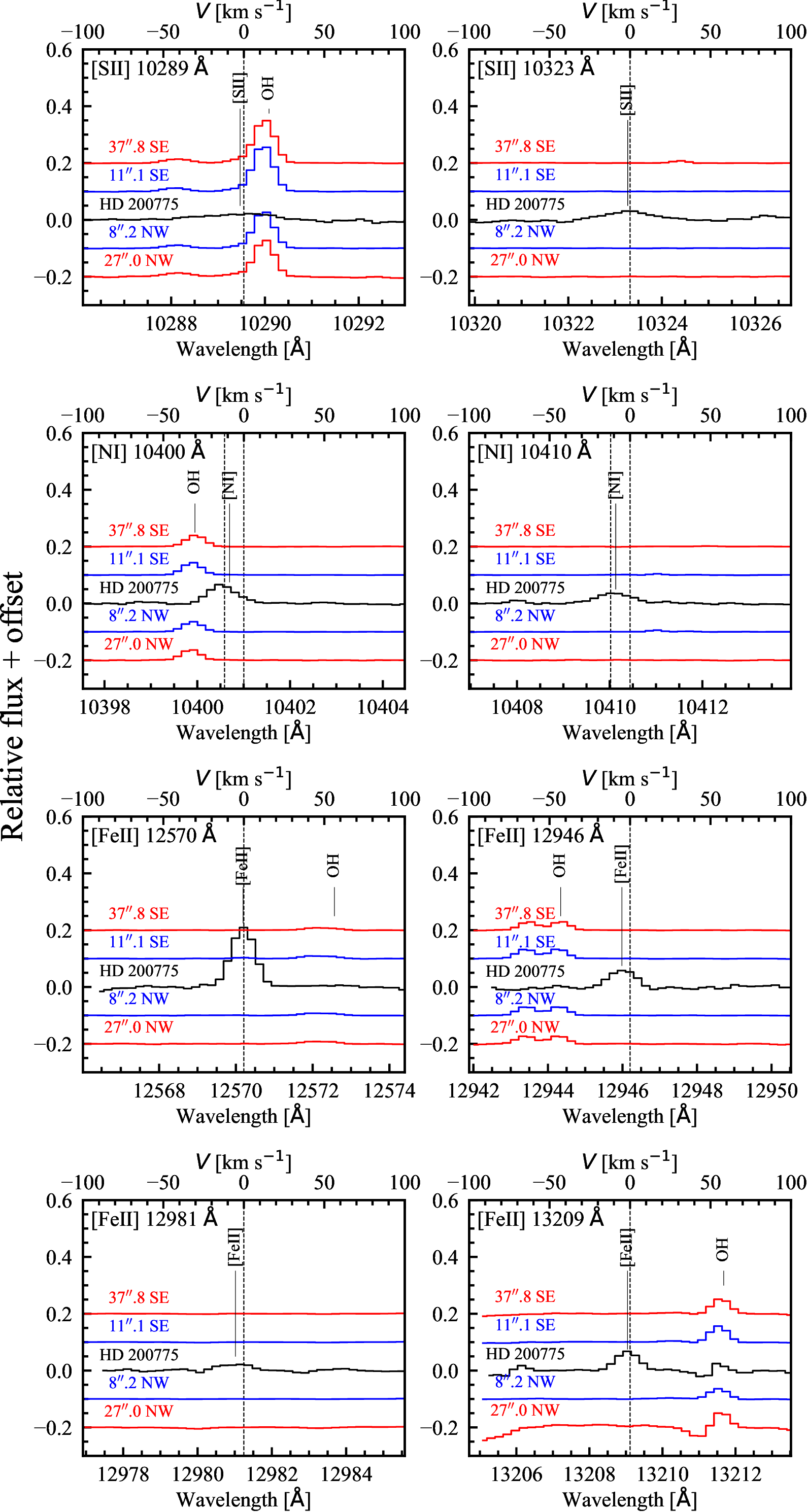}

\caption{Spectra of reference regions, surrounding HD 200775, and
  around the wavelengths of the forbidden lines (colored lines).
The intensities of the reference spectra are corrected
to be comparable to the continuum line of HD 200775.
For reference, the spectra of HD 200775 are shown by black lines.
The locations of the detected forbidden lines are indicated by the
respective line names.
OH emission lines of terrestrial atmospheric origin \citep{Oliva+2013}
 are labeled ``OH'' in the figure.}
\label{fig:compare}
  \end{center}
\end{figure*}

In this observation, a long slit of length $\sim$45$\arcsec$ directed
toward ${\rm PA} = 167^\circ$ was used.
Using 2D spectra reduced in the procedure of
Section~\ref{subsec:reduction} but without sky subtraction and
telluric corrections, we extracted the spectra of two regions
approximately 10$\arcsec$ and 30$\arcsec$ from the optical center of
HD 200775, with a width of $\sim$$7\farcs5$.
Because two reference regions can be set for each slit position A and
B,
there are four reference regions in total: 27$\farcs$0 northwest,
8$\farcs$2 northwest, 11$\farcs$1 southeast, and 37$\farcs$8 southeast
of HD 200775.
Figure~\ref{fig:ref_region_map} shows the spatial map of the reference
regions around HD 200775.
Because the focal plane array of WINERED is affected by latency and
leaves afterimages for several hours when bright objects are taken for
long periods \citep{Ikeda2022}, we selected the regions unaffected by
afterimages as reference regions.
The extracted spectra of the reference regions around the wavelengths of
the \SII, \NI, and \FeII{}
are shown with colored lines in Figure~\ref{fig:compare}.
The reference spectra are relative fluxes to the HD200775 spectra
normalized for the continuum level.
For comparison, the HD 200775 spectra, which are reduced in the
procedure in Section~\ref{subsec:reduction} (including sky subtraction
and telluric correction), are also shown (black lines). 
The positions of OH emission lines of a terrestrial atmospheric
origin \citep{Oliva+2013} are marked in the figure.

In the reference spectra, there are no positive features at wavelengths
within 50 km s$^{-1}$ of the forbidden lines detected in HD 200775
except around \SII{} 10289 \AA, \NI{} 10400 \AA, and \FeII{} 12946 \AA.
Therefore, all lines except these three forbidden lines cannot arise
from contamination from the reference regions or OH emission lines.
All features around the \SII{} 10289 \AA, \NI{} 10400 \AA, and \FeII{}
12946 \AA{} lines seen in the reference spectra are consistent with
the wavelengths of the OH emission lines and are of similar intensity
to all reference spectra.
Therefore, the positive features in the reference spectra are probably
due to OH emission lines rather than the forbidden lines
detected in HD 200775, which have somewhat different velocities.
We cannot rule out that the forbidden lines with almost identical
velocities and intensities are coincidentally observed in all four
reference regions around HD 200775; however, in this case, the
features are detected at wavelengths near the other forbidden lines of
the same element because the three lines are multiplets.
(For example, if \SII{} is detected at 10289 \AA{}, then \SII{} will
also be detected at 10323 \AA{}.
The same is true for \NI{} 10410 \AA{} for \NI{} 10400 \AA{} and 
\FeII{} 12570/129811/13209 \AA{} for \FeII{} 12946 \AA.)
In summary, it is highly unlikely that the forbidden lines detected in
HD 200775 represent contamination from surrounding regions.
In addition, the velocities of the peaks of the forbidden lines in HD
200775 differ from those of the positive features seen in the
reference spectra ($\gtrsim$ 20 km s$^{-1}$), so the forbidden lines
(including the \SII{} 10289 \AA, \NI{} 10400 \AA, and \FeII{} 12946
\AA\ forbidden lines) are unlikely to be OH emission lines.

 \subsection{Disk Winds around Intermediate-Mass Star}
 \label{subsec:imstarpe} 

Blueshifted forbidden emission lines, such as \OI{}, \SII{}, and
\NeII{}, are known as disk wind tracers \citep[e.g.,][]{Pascucci2023}.
Although the targets are mostly limited to low-mass stars,
\citet{Acke+2005} presented high spectral resolution optical spectra
of 49 Herbig Ae/Be stars in a search for the \OI{} 6300 \AA{} line.
The vast majority of stars in their sample showed emission
lines with $\Delta V_{\rm FWHM} < 100$ km s$^{-1}$.
Based on the profiles and the 5577/6300 \OI{} ratio, \citet{Acke+2005}
suggested that the lines originated in the surface layers of the
protoplanetary disks rather than in the disk winds as suggested for
low-mass stars. The profiles might be signatures of Keplerian
rotation.
The authors' interpretation was supported by the absence of the \SII{}
6731 \AA{} line in the emissions from their sample.
The profiles of a small fraction of their sample, including HD 200775,
show a high-velocity blue wing that cannot be explained by a Keplerian
disk but is plausibly due to outflow.
The emission lines of such objects, including HD 200775, usually show
low-velocity double peaks on the blueshift and redshift sides of the
stellar RV center.
Because their profiles were consistent with the expected profiles of
the emission-line regions on the disk surface, \citet{Acke+2005} again
interpreted the observed \OI{} lines as disk emission features coming
from the disk's atmosphere.

In the present paper, all detected forbidden lines from HD 200775 show
LVCs with blueshifted features.
Such features characterize the forbidden lines considered as disk wind
tracers of low-mass stars.
The morphology of the lines detected here differs from the \OI{}
lines (double-peaked LVCs and HVC) observed in HD 200775 by
\citet{Acke+2005}.
Meanwhile, \citet{Acke+2005} detected no \SII{} 6731 \AA{} lines from
HAeBes, which the authors interpreted as the absence of disk winds.
In contrast, our observations detected \SII{} 10289/10323 \AA{}\ lines
in HD 200775.
These lines are transitions from a common upper level with \SII{} 4068
\AA{}, which is considered a disk wind tracer.
Again, the results suggest disk wind in HD 200775 and can be traced
through lines detected in our observations, 
although the origin of these lines must be investigated more precisely
(e.g., by model fitting) in future work.
HD 200775 is a massive intermediate-mass object, and the mass is
located in the boundary region between intermediate- and high-mass
stars.
To our knowledge, we report the first indication of disk winds in such
a massive intermediate-mass star from forbidden lines.
The results suggest that disk winds are a universal process during
disk evolution not only in low-mass stars but also in intermediate-
and possibly high-mass stars.

The optically thick MHD disk winds are suggested to be effective in
the early stages of the evolutionary process of disk dissipation
(e.g., \citealt{Suzuki2009}).
Theoretical studies (e.g., \citealt{Hollenbach&Gorti2009},
\citealt{Ercolano&Owen2010}) proposed that the inner winds in the
early stages block energetic radiation from the central star and
surrounding regions, preventing energetic radiation from reaching the
disk and promoting photoevaporation, which is supported by the results
of subsequent observations of \NeII{} forbidden lines
\citep{Pascucci2020}.
Photoevaporation then becomes effective in the middle stage of
evolution, when the stellar and disk winds are weakening.
The strength of stellar and disk winds is quantified by the mass-loss
rate ($\dot{M}_{\rm loss}$), which correlates with the mass accretion
rate to the central stars ($\dot{M}_{\rm acc}$).
In theoretical models, photoevaporation is only effective when
$\dot{M}_{\rm acc} \sim 10^{-7}$ to $\dot{M}_{\rm acc} \sim 10^{-9}$
$M_\odot$ yr$^{-1}$ (e.g., \citealt{{Wang2017ApJ},{Nakatani2018},
  {Picogna2019}}).
Observationally, it has been proposed that photoevaporation begins to
take effect when the mass accretion rate falls below $10^{-8}$
$M_\odot$ yr$^{-1}$ \citep{Pascucci2020}.
HD 200775 is very young ($\sim$0.1 Myr old; \citealt{Alecian+2008})
and is estimated to have a high mass accretion rate ($\dot{M}_{\rm
    acc} = 10^{-4.2}$ $M_\odot$ yr$^{-1}$; \citealt{Arun+2019}) and a
high mass-loss rate ($\dot{M}_{\rm loss} = 10^{-6.55}$ $M_\odot$
yr$^{-1}$; \citealt{Damiani+1994}).
Therefore, it seems natural to assume that this is an MHD disk disk
wind. However, the mass-loss rate due to photoevaporation is expected
to be high for intermediate-mass stars of 10 $M_\odot$, about 100
times higher than for low-mass stars of 1 $M_\odot$
\citep{Komaki2021}.

Recently, \citet{Kunitomo+2021} calculated the time evolution of the
photoevaporation rates around young stars in the mass range covering
intermediate-mass stars (0.5-5 $M_\odot$ stars).
After considering the effect of stellar evolution on the evolution of
protoplanetary disks, 
they suggested that far-UV emissions from the central star are
sufficiently high for the generation of effective photoevaporative
winds by a 5 $M_\odot$ intermediate-mass star at a very young stage
($\lesssim$10$^4$ yr).
Therefore, photoevaporation is also a possible process in HD 200775,
although HD 200775 is heavier than the upper mass considered in
Kunitomo et al.'s model.
However, the effects of the MHD disk wind and the stellar wind were
excluded from this model and instead marked for future work.

\subsection{NIR \NI{} Lines as New Disk Wind Tracers}
 \label{subsec:new_tracers} 

Forbidden lines, such as \OI{} 5577/6300 \AA{} and \SII{} 4068/4076
\AA{} in the optical band and \NeII{} 12.8 $\mu$m in the MIR
wavelength region, are recognized as disk wind tracers (e.g.,
\citealt{Pascucci2020}). 
The forbidden lines detected here, \NI{}, \FeII{}, and \SII{}, are all
in the NIR range, which can be observed with high sensitivity.
This is a significant advantage over \NeII{} 12.8 $\mu$m in the
MIR-wavelength range as MIR wavelengths are not easily detected in
ground-based observations, because they are largely affected by 
Earth's atmosphere and the thermal ambient radiation.
The lines in the NIR range also compare favorably with \OI{} and
\SII{} lines in the optical range.
In general, objects in the very young phase that are still embedded in
molecular clouds are difficult to detect due to their large extinction
in the optical bands.
The extinction for \NI{} 10400/10410 \AA{} ($A_y /A_V \simeq 0.4$) is
significantly smaller than the extinction for \SII{} 4068/4076 \AA{}
($A_g /A_V \simeq 1.2$) and \OI{} 5577/6300 \AA{} ($A_r /A_V \simeq
0.8$) \citep{Wang2019}.
Using recently developed high-sensitivity NIR high-dispersion
spectrographs, high-S/N spectra can be obtained in a short time.
Therefore, \NI{}, \FeII{}, and \SII{} lines are very useful for
evaluating the presence or absence of disk winds and their physical
parameters.

The \NI{} lines also have an advantage in that they are relatively
easy to detect. 
While \NeII{} 12.8 $\mu$m, \OI{} 5577 \AA{}, and
\OI{} 6300 \AA{}
are limited to the observed regions of electron density in the range
$N_e=10^2$--$10^6$ cm$^{-3}$ (for $T_e = 2000$--10,000 K),
$N_e=10^6$--$10^7$ cm$^{-3}$
(for $T_e = 4000$--10,000 K), and
$N_e=10^5$--$10^7$ cm$^{-3}$
(for $T_e = 5000$--8000 K), respectively
(Figure~5 of \citealt{Pascucci2020}),
the \NI{} 10400/10410 \AA{} quadruplet lines are predicted to be the
strongest among the different \NI{} lines in the wide range of
electron densities from $N_e=10^2$--$10^{10}$ cm$^{-3}$ for $T_e =
5,000$--10,000 K according to the calculations by
\citet{Kastner1997}.
In addition, the total intensity of the four emission lines is
predicted to be very sensitive to electron density. This intensity
dependence over the wide electron density of the \NI{} lines enables
compensative studies of disk wind at any evolutionary stage.
Actually, in Section~4.3 we suggested that disk winds could occur in
HD 20075, which is still very young ($\sim$0.1 Myr old;
\citealt{Alecian+2008}) and has a very high mass accretion rate
($>$$10^{-5}$ $M_\odot$ yr$^{-1}$; \citealt{Arun+2019}).
This indicates that the NIR lines, including the \NI{} lines, are very
sensitive to disk winds even at a very early stage.
This is in contrast to \NeII, which shows signs only at the late stage
of disk evolution \citep{Pascucci2009}.
The NIR lines can serve as important tracers for capturing signs of
disk winds in the early phases of protoplanetary disks, when material
is still abundant.

The utility of \NI{} 10400 and 10410 \AA{} may be further enhanced by
their quadruple line structure, which consists of two overlapping
doublets (10400.587 and 10410.021, and 10401.004 and 10410.439
\AA{}). The intensity ratio of the double lines formed by the
separation of lower levels is determined solely by the ratio of the
Einstein A coefficients, but the intensity ratio of the double lines
formed by the separation of upper levels is known to depend strongly
on the electron temperature and density
\citep[e.g.,][]{Osterbrock2006}.
Especially when the energy difference between upper levels is small,
as is the case with \NI{} 10400 and 10410, the intensity ratio is
dominated by collisional deexcitation, making it very sensitive to
electron density. Therefore, the emission lines of \NI{} can
potentially be a powerful probe for investigating the electron density
of the disk wind. Furthermore, by using two pairs of double lines in
\NI{} simultaneously, the accuracy of the electron density
determination can be improved. In the future, we plan to develop
photoionization models and advance spectroscopic diagnostics for
\NI{}. This will clarify the physical properties of the disk wind. In
addition to the \NI{} line, the \FeII{} and \SII{} lines are
multiplets and can be used to derive physical parameters in the same
way as \NI{}.


\section{Conclusion} \label{sec:Conclusion}

In this study, we focused on intermediate-mass stars to search for new
disk wind tracers (MHD disk wind and/or photoevaporative wind) in the
NIR-wavelength range.
We performed high-resolution NIR spectroscopy ($R=28$,000; $\lambda =
0.91$--1.33 $\mu$m) on an intermediate-mass star, HD 200775, using
WINERED.
HD 200775 is a massive intermediate-mass star that has protoplanetary
disks (a binary star with a mass of approximately 10 $M_\odot$ each)
and is very young ($\sim$0.1 Myr old). The main results are summarized
as follows:

\begin{enumerate}
 \item In the obtained HD 200775 spectra, eight forbidden lines that
   can be disk wind tracers of three elements were detected:
       two in \SII{} (10289 and 10323 \AA), two in \NI{} (10400 and
       10410 \AA), and four in \FeII{} (12570, 12946, 12981,
      and 13209 \AA). 
       These are all multiple lines. Each of the \NI{} lines consists
       of doublet lines, making a total of quadruple lines.
      This is the first time, to our knowledge, that \NI{} lines were
      detected in YSO with a doublet deblended (10400 and 10410 \AA).

 \item We performed Gaussian fitting on the obtained spectra to estimate
       the central velocities and the velocity widths of the detected
       forbidden lines.
       In all the forbidden lines, the central velocity showed
       slightly blueshifted features of $\gtrsim$$-$5 km s$^{-1}$.
       The FWHMs of \NI{} and \FeII{} were narrow ($\sim$15--20 km
       s$^{-1}$), whereas those of \SII{} were slightly wider
       ($\sim$30--40 km s$^{-1}$).
       All the forbidden lines were classified as
         LVCs, suggesting a disk wind origin. 
       Based on the velocity fields of these emission lines, the
       \NI{} and \FeII{} lines were categorized as LVC-NC, while
       the \SII{} lines correspond to the boundary between LVC-NC and
       LVC-BC.
       These lines have been assumed as signatures of the
       jets of low-mass YSOs because they had been detected with various
       components (including HVCs) in such stars.
       The detection of all lines as LVCs in HD 200775 but as various
       components in low-mass T Tauri stars may be attributed to
       differences in the masses of the YSOs.

 \item HD 200775 is located around the center of a reflection nebula,
   NGC 7023, which is associated with PDRs, and the detected forbidden
   lines could be attributed to radiation from these regions.
       The detected lines can also be a result of contamination from
       OH emission lines during the observation.
       To verify these possibilities, we analyzed the spectra of the
       surrounding regions of HD 200775 without sky subtractions and
       telluric corrections.
       The reference spectra showed positive features around three of
       the eight forbidden lines detected in HD 200775, which were
       probably OH lines but were $\gtrsim$20 km s$^{-1}$ away from
       the forbidden lines in velocity.
       This suggests that the forbidden lines are not a result of
       contamination from the associated regions (i.e., the nebula and
       PDRs) or OH emission lines but are intrinsic features of the
       object.

     \item The object in which the disk wind was observed herein, HD
       200775, is an intermediate-mass star whose mass is in the
       boundary region with high-mass stars.
       This is the first indication of disk winds in such massive
       intermediate-mass stars from forbidden lines. 
       This suggests that disk winds are a universal process in disk
       evolution not only for low-mass but also for intermediate-mass
       stars and possibly for high-mass stars.
       Recent observational results suggest that photoevaporative wind
       is ineffective in the early stages when the accretion rate is
       high because mass flows block the radiation from the central
       star, suggesting that the MHD disk wind is dominant, while the
       photoevaporative wind may be active in intermediate-mass stars
       because of the presence of strong far-UV in the early stages.
       
 \item All the forbidden lines detected herein are in the NIR region;
   thus, they have a higher sensitivity for observing young stars from
   very early stages compared to previous disk wind tracers, such as
   \OI{} and \SII{} in the optical region and \NeII{} in the MIR
   region.
       Furthermore, all the forbidden lines are multiplets, and from
       their intensity ratio the physical parameters (temperature and
       density) of the gas can be determined by developing
       photoionization models.
       In particular, the \NI{} lines, whose energy difference between
       the upper layers is very small, can be quite promising for the
       derivation with very high accuracy.
      This could reveal the physical properties of the disk wind gas.
      However, the discussion herein is limited to one object (HD
      200775); thus, further investigation is needed to determine
      whether it can be applied to other YSOs.
       
\end{enumerate}

\begin{acknowledgments}
  We thank Dr. Masanobu Kunitomo, Dr. Riouhei Nakatani, and Dr. Kei
  Tanaka for many useful comments.
  WINERED was developed by the University of Tokyo and the Laboratory of
Infrared High-resolution Spectroscopy, Kyoto Sangyo University, under
the financial support of KAKENHI (Nos. 16684001, 20340042, and 21840052)
and the MEXT Supported Program for the Strategic Research Foundation at
Private Universities (Nos. S0801061 and S1411028).
This study was partly supported by Koyama Space Science Institute of
Kyoto Sangyo University.
KH is supported by JST SPRING, Grant Number JPMJSP2157.

 {\it Software:} IRAF \citep{Tody1993}, WINERED pipeline \citep{Hamano2024}. 

\end{acknowledgments}
\vspace{1em}

\bibliography{katoh2022}{}
\bibliographystyle{aasjournal}




\begin{deluxetable*}{cccccc}
  \tablecaption{Summary of WINERED Observations\label{tab:ObsSummary}}
  \tablewidth{700pt}
  \tabletypesize{\scriptsize}
  \tablehead{
  \colhead{Object} & \colhead{Spectral Type} & \colhead{Object Type} &
  \colhead{Obs Date (UT)} & \colhead{$t_{\rm{exp}}$} &
  \colhead{Airmass} \\
  \colhead{} & \colhead{} & \colhead{} & \colhead{(yyyy-mm-dd hh:mm--hh:mm)} & \colhead{(s)} & \colhead{($\sec z$)}
  }
  \startdata
  HD 200775 & B3 & Target & 2015-07-20 15:33--16:19 & 2400 & 1.2 \\
  $\rho$ Peg & A1V & Telluric standard & 2015-07-20 16:49--17:01 & 480 & 1.2 \\
  \enddata
\end{deluxetable*}

\begin{deluxetable*}{lccccccccc}
 \tablecaption{Properties of Forbidden Lines Detected in HD
 200775 \label{tab:flinelist}}
  \tablewidth{700pt}
  \tabletypesize{\scriptsize}
  \tablehead{
  \colhead{Line} & \colhead{$\lambda_{\rm vac}$} &
  \colhead{$\lambda_{\rm air}$} & \colhead{$A_{\rm ul}$} &
  \colhead{Transition} & \colhead{S/R} &
  \colhead{$V_c$} &
  \colhead{$\Delta V_{\rm FWHM}$} &
  \colhead{$I_{\rm peak}$} &\\
  \colhead{} & \colhead{(\AA)} & \colhead{(\AA)} & \colhead{(s$^{-1}$)} &
 \colhead{} & \colhead{} & \colhead{(km s$^{-1}$)} & \colhead{(km s$^{-1}$)}
 & \colhead{} \\
\colhead{(1)} & \colhead{(2)} &
  \colhead{(3)} & \colhead{(4)} &
  \colhead{(5)} & \colhead{(6)} &
  \colhead{(7)} & \colhead{(8)} &
  \colhead{(9)} 
  } 
  \startdata
  \SII{} 10289 & 10289.55 & 10286.73 & 1.15E-01 & 3s$^2$\ 3p$^3$\ $^2
 $D$^o_{3/2}$--3s$^2$\ 3p$^3$\ $^2$P$^o_{3/2}$ & 12.5 &
  $-$2.3 $\pm$ 2.3 & 43.2 $\pm$ 5.1 & 0.020 $\pm$ 0.002 \\
 \SII{} 10323 & 10323.32 & 10320.49 & 1.57E-01 & 3s$^2$\ 3p$^3$\ $^2$D$^o_{5/2}$--3s$^2$\ 3p$^3$\ $^2$P$^o_{3/2}$ & 20.8 &
  $-$1.5 $\pm$ 1.7 & 31.3 $\pm$ 3.5 & 0.030 $\pm$ 0.003 \\
 %
 \multicolumn{1}{l|}{\multirow{2}{*}{\NI{} 10400}} & 10400.59 & 10397.74 & 6.10E-02 & 2s$^2$\ 2p$^3$\ $^2$D$^o_{5/2}$--2s$^2$\ 2p$^3$\ $^2$P$^o_{1/2}$ & \multirow{2}{*}{45.6} &
 \multirow{4}{*}{ $-$2.9 $\pm$ 1.1} & \multirow{4}{*}{12.3 $\pm$ 1.3} & 0.061 $\pm$ 0.003 & \\ 
 \multicolumn{1}{l|}{} & 10401.00 & 10398.16 & 3.45E-02 & 2s$^2$\ 2p$^3$\ $^2$D$^o_{5/2}$--2s$^2$\ 2p$^3$\ $^2$P$^o_{3/2}$ & &
  & & 0.013 $\pm$ 0.002 \\
 \multicolumn{1}{l|}{\multirow{2}{*}{\NI{} 10410}} & 10410.02 & 10407.16 & 2.64E-02 & 2s$^2$\ 2p$^3$\ $^2$D$^o_{3/2}$--2s$^2$\ 2p$^3$\ $^2$P$^o_{3/2}$ & \multirow{2}{*}{17.3} &
 & & 0.026 $\pm$ 0.001 \\
 \multicolumn{1}{l|}{} & 10410.44 & 10407.59 & 5.31E-02 & 2s$^2$\ 2p$^3$\ $^2$D$^o_{3/2}$--2s$^2$\ 2p$^3$\ $^2$P$^o_{1/2}$ & &
 & & 0.020 $\pm$ 0.003 \\
 \FeII{} 12570 & 12570.21 & 12566.77 & 4.74E-03 & 3d$^6\left(^5\rm{D}\right)$\ 4s\ a\ $^6$D$_{9/2}$--3d$^6\left(^5\rm{D}\right)$\ 4s\ a\ $^4$D$_{7/2}$ & 53.0 &
 $-$0.5 $\pm$ 1.0 & 12.1 $\pm$ 0.8 & 0.207 $\pm$ 0.006 \\
 \FeII{} 12946 & 12946.20 & 12942.66 & 1.98E-03 & 3d$^6\left(^5\rm{D}\right)$\ 4s\ a\ $^6$D$_{5/2}$--3d$^6\left(^5\rm{D}\right)$\ 4s\ a\ $^4$D$_{5/2}$ & 24.7 &
  $-$5.0 $\pm$ 1.1 & 14.0 $\pm$ 1.3 & 0.063 $\pm$ 0.003 \\
 \FeII{} 12981 & 12981.25 & 12977.70 & 1.08E-03 & 3d$^6\left(^5\rm{D}\right)$\ 4s\ a\ $^6$D$_{1/2}$--3d$^6\left(^5\rm{D}\right)$\ 4s\ a\ $^4$D$_{3/2}$ & 11.2 &
  $-$5.2 $\pm$ 1.6 & 18.3 $\pm$ 3.3 & 0.022 $\pm$ 0.003 \\
 \FeII{} 13209 & 13209.11 & 13025.50 & 1.31E-03 & 3d$^6\left(^5\rm{D}\right)$\ 4s\ a\ $^6$D$_{7/2}$--3d$^6\left(^5\rm{D}\right)$\ 4s\ a\ $^4$D$_{7/2}$ & 10.7 &
  $-$1.5 $\pm$ 1.8 & 12.1 $\pm$ 4.9 & 0.069 $\pm$ 0.013 
  \enddata

\tablecomments{Column (2): wavelength in a vacuum.
Column (3): wavelength in the air.
Column (4): Einstein $A$ coefficient.
 Column (6): S/N.
 Column (7): centroid velocity. 
 Column (8): velocity width in FWHM.
Column (9): peak intensity relative to the continuum level.}
\end{deluxetable*}

\end{document}